\documentclass{Interspeech}

\interspeechcameraready

\usepackage[utf8]{inputenc}
\usepackage[T1]{fontenc}
\usepackage{babel}
\usepackage{multirow} %
\usepackage{hhline} %
\usepackage{subcaption}
\usepackage{hyperref}
\usepackage{seqsplit}
\usepackage{ragged2e}
\usepackage{etoolbox}
\usepackage{enumitem} 
\usepackage{multirow}
\usepackage{float}
\usepackage{placeins}
\usepackage{makecell}

\usepackage{eso-pic}
\usepackage{xcolor}
\usepackage{tikz}

\AddToShipoutPictureBG*{%
  \AtPageLowerLeft{%
    \raisebox{1cm}{%
      \makebox[\paperwidth]{%
        \hfill
        \tikz[baseline]\node[inner sep=2pt,fill=white,opacity=0.7,text opacity=1,rounded corners]
          {\small\itshape This paper is accepted (camera‐ready) for Interspeech 2025};
        \hspace{1cm}%
      }%
    }%
  }%
}

\title{Evaluating the Usefulness of Non-Diagnostic Speech Data for Developing Parkinson's Disease Classifiers}

\author{Terry Yi}{Zhong}
\author{Esther}{Janse}
\author{Cristian}{Tejedor-Garcia}
\author{Louis ten}{Bosch}
\author{Martha}{Larson}

\affiliation[nocounter]{Centre for Language Studies}{Radboud University}{the Netherlands}
\email{yi.zhong, esther.janse, l.tenbosch, cristian.tejedorgarcia, martha.larson - @ru.nl}

\begin{document}

\maketitle


\raggedbottom

\begin{abstract}

Speech-based Parkinson's disease (PD) detection has gained attention for its automated, cost-effective, and non-intrusive nature. As research studies usually rely on data from diagnostic-oriented speech tasks, this work explores the feasibility of diagnosing PD on the basis of speech data not originally intended for diagnostic purposes, using the Turn-Taking (TT) dataset. Our findings indicate that TT can be as useful as diagnostic-oriented PD datasets like PC-GITA. We also investigate which specific dataset characteristics impact PD classification performance. The results show that concatenating audio recordings and balancing participants' gender and status distributions can be beneficial. Cross-dataset evaluation reveals that models trained on PC-GITA generalize poorly to TT, whereas models trained on TT perform better on PC-GITA. Furthermore, we provide insights into the high variability across folds, which is mainly due to large differences in individual speaker performance. 
\end{abstract}


\section{Introduction}




Parkinson's disease (PD) is the second most common neurodegenerative disorder, affecting over 10 million people worldwide~\cite{1_ngo2022computerized, poewe2017parkinson}. During the prodromal stages of PD, patients may develop various speech problems up to five years before the onset of significant motor impairments~\cite{pdspeech1pinto2004treatments, pdspeech2rusz2013imprecise, mu2017parkinson}.
Speech-based PD detection has attracted attention due to its automated, cost-effective, and non-invasive approach~\cite{biomarkertelho2024speech, SSL4PR, Foundation4PDpurohitautomatic, PCGITA2oliveiraPilotStudySpeech2025, vasquez-correaConvolutionalNeuralNetworks2019}.
However, these studies usually rely on speech data collected from diagnostic-oriented speech tasks performed by participants. 

Relaxing the need for particular speech tasks is advantageous in several ways. First, it saves time and costs for both participants and healthcare systems by reducing extra clinical visits.\footnote{\scriptsize\url{https://doi.org/10.48550/arXiv.2501.03536}} Second, it facilitates faster referrals and follow-ups, especially in resource-limited settings with restricted healthcare access~\cite{referralsprabhod2024role}. 
Third, the "observer effect"~\cite{observerSVENSBERG20212136} may be more pronounced on articulation in particular tasks like diadochokinesis (DDK),\footnote{\scriptsize{Repetition of syllable sequences: /pa-ta-ka/, /pe-ta-ka/, /pa/, /ka/, /ta/, etc.}} which can undermine diagnostic reliability. Other tasks, like dialogue-based ones, may be more engaging, reducing participants' focus on optimizing their articulation performance.
Furthermore, no standardized speech assessment protocol exists at the PD diagnosis stage (i.e., with general practitioners or neurologists), making dedicated data collection for PD diagnosis challenging. 
Therefore, it would potentially be helpful, with the consent of the potential patients, to test for PD on speech data collected for other purposes.

A recent effort, ParkCeleb~\cite{ParkCeleb_favaroUnveilingEarlySigns2024}, has explored using speech data from non-clinical settings for PD detection, demonstrating the potential of utilizing celebrity speech from YouTube. 
Nevertheless, celebrity speech differs from typical populations, as patients are less accustomed to public speaking, making replication in clinical settings challenging. However, patients may have completed speech tasks for other purposes in healthcare contexts, offering alternative data sources.

Therefore, from a clinical point of view, it is more interesting to focus on situations in which speech is recorded in controlled settings, such as laboratories or clinical environments, for research or treatment purposes, but not specifically for PD diagnosis. 
For instance, the Turn-Taking (TT) dataset that we introduce in this work, originally developed for studying conversational turn-taking phenomena~\cite{wiemann2017turn} as a function of aging and in PD, differs from other PD corpora in the sense that the speech tasks are not "standardized" diagnostic-oriented speech tasks like DDK or sustained vowel. Rather, the dataset includes experimental question-answer paradigms to elicit speech from participants. 

In this work, we examine the potential of speech data collected in non-diagnostic, controlled settings (using the TT dataset) for developing a PD classifier. Additionally, we explore whether dialogue-oriented speech elicitation provides clinically relevant diagnostic insights. We analyze key characteristics of the dataset, including recording length, gender balance, and disease status balance, to assess their impact on performance.
Motivated by two recent studies~\cite{SSL4PR, Foundation4PDpurohitautomatic} that reported high standard deviations (std) for classification performance across test folds without discussing the underlying causes, we analyze factors contributing to this variability.
To the best of our knowledge, this is also the first study to explore PD detection using a Dutch speech-based dataset (TT). Therefore, the research questions (\textit{RQ}s) of our work are:


 \noindent \textit{RQ}1:\makebox[0.15cm][l]{ }Can controlled setting speech data (designed for non-diagnostic purposes) be useful for developing a PD classifier?


\noindent \textit{RQ}2:\makebox[0.15cm][l]{ }Which dataset characteristics affect within-dataset classification performance and cross-dataset generalization?

 \noindent \textit{RQ}3:\makebox[0.15cm][l]{ }In case of large variability in classification performance across test folds, which factors contribute to this variability?

\section{Datasets}
\label{sec:levels}
\subsection{
Different Levels of Diagnostic-Purpose}

We start by clarifying our definition of different levels of PD diagnostic-purpose, as the distinction between different datasets in terms of purpose may not be immediately clear. Here we only consider publicly available speech datasets~\cite{LisannevangelderenInnovativeSpeechBasedDeep2024} containing PD labels. (1) \textbf{High-level} diagnostic-purpose datasets, such as PC-GITA~\cite{PCGITAorozco2014new} and NeuroVoz~\cite{Neurovozmendes2024neurovoz}, are primarily designed for PD diagnosis, as demonstrated by their use in training PD classification models in their original study. These datasets often include tasks commonly used in PD diagnosis, like the DDK tasks. 
They usually include tests for quantifying PD severity, such as the Unified PD Rating Scale (UPDRS) and the Hoehn and Yahr (H\&Y) Scale. Constructing such datasets requires significant effort and costs due to the inclusion of these specialized tasks. 
In contrast, (2) \textbf{Mid-level} datasets are those recorded in controlled environments (e.g., laboratories, clinics) for research or treatment purposes but were not specifically designed for PD diagnosis, like the TT dataset. These datasets may not include widely used PD severity scales like UPDRS and often feature varying tasks, including less explored ones, such as dialogue-based tasks that may be more engaging for patients than traditional DDK or reading tasks.
Finally, (3) \textbf{Low-level} diagnostic-purpose datasets like ParkCeleb~\cite{ParkCeleb_favaroUnveilingEarlySigns2024} are obtained from in-the-wild speech data without any specific recording design. These datasets lack research objectives and tend to suffer from high background noise and complex multi-speaker environments. Furthermore, replicating these conditions with new participants is usually difficult, making it challenging to generalize the dataset for clinical use.

To the best of our knowledge, no prior study has explored the diagnostic potential of mid-level diagnostic-purpose PD datasets, such as the TT dataset. However, understanding the value of these datasets is crucial, as they offer the potential to allow the community to leverage a broader range of data.

\subsection{TT Dataset (Mid-Level Diagnostic-Purpose)}

This new Dutch speech-based PD dataset was collected at Radboud University during 2023 and 2024, using a Dell Latitude 5590 laptop, a MOTU M2 audio interface, and a Sennheiser K3N/ME 40 microphone. Sound playback was performed via two Logitech Z130 speakers. 
The inclusion criteria were:
(1) native Dutch speakers capable of completing all tasks;
(2) healthy controls (HC) aged 60–85 years; and
(3) PD participants aged 18–85 years with a diagnosis of Parkinson's disease confirmed by a neurosurgeon.
For inclusion in the present study, we only included participants who provided their informed consent that their data could be used for future research by internal and external researchers.\footnote{\scriptsize{The dataset will be made available to the research community soon.}}

\subsubsection{Tasks and Conditions}
The TT dataset consists of three components: two experimental dialogue-oriented paradigms that elicit short responses (either single nouns or short sentences) and a picture naming task~\cite{pictureglaser1992picture}. The dialogue tasks were designed to explore whether PD affects the ability to plan speech while listening and to examine the timing of turn-taking in PD, particularly in response to cues signaling turn finality. In the first dialogue experiment (EarlyLate, based on~\cite{earlylatebogels2018planning}), participants see a four-picture display on the screen and hear a prerecorded question (e.g., "\textit{Which object...?}"). The phrasing of the question is designed so that the critical information to identify the correct object appears either early or late in the question, giving participants varying amounts of time to prepare their responses. In the second dialogue experiment (BoundaryTone, based on~\cite{boundarytonebarthel2017next}), participants see a set of objects on the screen and are told that their confederate has fewer objects. After the confederate provides a prerecorded description of their objects, participants must complete the list by identifying additional objects they see (e.g., "\textit{I also have an X and a Y}"). The critical manipulation in the confederate's speech involves cues to turn-finality, such as the use of "and" before the last object mentioned and the presence of a clear utterance-final boundary tone. The third component is a picture-naming task, where participants view a single object and are instructed to name it quickly.

\subsubsection{Data Preprocessing}
All audio recordings were converted to a single channel and peak-normalized to ensure consistent volume levels. We applied Silero\footnote{\scriptsize\url{https://github.com/snakers4/silero-vad}} Voice Activity Detection to remove silence at the beginning and end of each recording, retaining only the meaningful speech segments. To maintain consistent recording durations, we calculated the mean length across all samples in each task and removed recordings that deviated more than three standard deviations from the mean. Since some recordings might contain instructor voices, we used pyannote speaker diarization 3.1\footnote{\scriptsize\url{https://huggingface.co/pyannote}} to detect multi-speaker segments. After manually verified by human listening, recordings confirmed to contain multiple speakers were excluded.\footnote{\scriptsize{\url{https://github.com/terryyizhongru/TurnTakingPD}}}

\subsection{PC-GITA Dataset (High-Level Diagnostic-Purpose)}

As a speech corpus specifically designed for PD diagnosis, we selected PC-GITA~\cite{PCGITAorozco2014new}. Since its release, it has been extensively used as a dataset for developing and validating PD classification methods~\cite{SSL4PR, PCGITA2oliveiraPilotStudySpeech2025, PCGITA3gallo-aristizabalAutomaticClassificationParkinsons2024a, ParkCeleb_favaroUnveilingEarlySigns2024, PCGITAorozco2014new}.
This dataset includes 100 participants: 50 PD patients and 50 HC. Each group has 25 male and 25 female participants. All PD patients were diagnosed by a neurologist, while HC participants had no PD symptoms or other neurodegenerative disorders. 

All recordings were made when patients were in their "on" medication state. The recordings were collected in Colombian Spanish in a soundproof booth. All audio was recorded at a 44.1 KHz sampling rate with 16-bit resolution. The speech tasks from PC-GITA used in this work include DDK tasks, read text, sentences, and monologues, as in~\cite{SSL4PR}.

\subsection{Datasets Differences}\label{sec:diff}
We compare the PC-GITA and TT datasets across several aspects to obtain a general view of the differences between them and offer insights into improving their comparability for subsequent experiments.

\vspace{-0.15cm}

\begin{table}[ht!]
\centering
\footnotesize 
\caption{Demographic information for HC and PD groups across the PC-GITA, TT datasets, and TT-balanced subset.}
\vspace{-0.25cm}

\begin{tabular}{l@{}c@{}c@{}c@{}c@{}c}
\hline
\textbf{Dataset} & \textbf{Status } & \textbf{ Participants } & \textbf{ M/F } & \textbf{ Avg age (range) } & \textbf{ TAD (yrs)}\\ \hline
PC-GITA           & HC             & 50                & 25/25                & 60.98 (31-86)              \\
PC-GITA           & PD             & 50                & 25/25                & 61.02 (33-81)         & 11.2     \\ \hline
TT               & HC             & 33                & 16/17                & 71.16 (61-80)              \\ 
TT               & PD             & 38              & 23/15                & 67.97 (54-84)           & 7.5     \\ \hline
TT-balanced               & HC             & 30                & 15/15                & 71.03 (61-80)              \\ 
TT-balanced           & PD             & 30              & 15/15                & 68.27 (54-84)         & 7.6       \\ \hline
\end{tabular}
\label{tab:demographic_info}
\end{table}
\vspace{-0.15cm}

In Table~\ref{tab:demographic_info}, we compare the demographic information of the two datasets. The TT dataset contains fewer participants than PC-GITA, and the numbers across statuses and genders are not fully balanced. We selected a subset of TT to construct TT-balanced, ensuring balance in both gender and status, making it more comparable to PC-GITA. The average age in PC-GITA is slightly lower than in TT. Besides, the average time after diagnosis (TAD) in PC-GITA is longer than in TT, suggesting that PD patients in TT may be in an earlier stage of the disease.

In Table~\ref{tab:audio_stats}, we compare the basic audio characteristics of the two datasets. The sum of audio duration and average duration per participant is similar between the datasets. However, the mean duration of individual recordings in TT is significantly shorter than in PC-GITA, with a higher number of recordings per participant. To improve comparability with PC-GITA, we created TT-concat (the same demographic information as TT) by randomly concatenating four recordings per speaker of TT. For loudness, PC-GITA exhibits higher and more stable levels compared to TT, with a narrower range and less variability.  

\vspace{-0.15cm}

\begin{table}[ht!]
\centering
\footnotesize
\caption{Audio aspect statistics for PC-GITA, TT, and TT-concat set. Statistic values are shown as mean±std, (min, max). }
\vspace{-0.15cm}

\begin{tabular}{p{1.9cm}|p{1.4cm}|p{1.4cm}|p{1.4cm}}
\hline
\textbf{Audio aspect} & \textbf{PCGITA} & \textbf{TT} & \textbf{TT-concat} \\ \hline
Sum Duration(h) & 3.43 & 3.79 & 3.73 \\ \hline
Avg Dur/Subj(s) & 123.5 & 192.6 & 188.5 \\ \hline
Recordings/Subj & 18.0 & 142.7 & 35.3 \\ \hline
Duration statistics (s) & 6.9±12.2 (0.8,164) & 1.4±1.3 (0.3,19.3) & 5.3±2.6 (2.0,22.0) \\ \hline
Loudness statistics (dBFS) & –38±8.5 (–61,–12) & –45±10.8 (–90,–25) & –44±10.6 (–84,–27) \\ \hline
\end{tabular}
\label{tab:audio_stats}
\end{table}
\vspace{-0.1cm}

For the aspects related to PD, PC-GITA includes TAD and widely used PD-specific severity metrics, such as H\&Y and UPDRS. TT provides TAD and speech score of Radboud Oral Motor Inventory for PD (ROMP)~\cite{Rompkalf2011reproducibility}, which assesses self-reported speech problems. Speech tasks in PC-GITA are commonly used in PD diagnostic research, whereas TT includes specialized tasks designed for other research purposes.





\section{Method}

\subsection{Classification Model}
We utilized a machine learning classification approach, as motivated by recent advances~\cite{biomarkertelho2024speech,SSL4PR,gimeno2025unveiling}, to explore the potential of  PD diagnosis using TT.
Based on a literature review on classification performance using the PC-GITA dataset, we selected the classification model proposed in~\cite{SSL4PR} for our experiments, as it is the top-performing open-source method that employs nested speaker-independent cross-validation.

The model utilizes a pre-trained encoder to extract high-level speech features. A weighted sum with attention pooling aggregates frame-level representations and is then processed by two fully connected layers with ReLU activations. A final Sigmoid output layer generates the prediction. The model is trained end-to-end on the PD dataset, with all parameters updated jointly. We used the official code repository\footnote{\label{myfootnote}\scriptsize\url{https://github.com/K-STMLab/SSL4PR}} for implementation and selected WavLM Base\footnote{\scriptsize\url{https://huggingface.co/morenolq/SSL4PR-wavlm-base}} as the foundational encoder~\cite{chen2022wavlm}, which offers the best performance on PC-GITA.

\subsection{Experimental Setup}

The same evaluation metrics as in~\cite{SSL4PR} were used, including accuracy, precision, F1-score, AUC-ROC,\footnote{\scriptsize{We corrected an error in the computation of the AUC-ROC in the code of~\cite{SSL4PR}.}} sensitivity, and specificity. 
For the PC-GITA, we used the provided training and test splits.\footnotemark[7] We conducted experiments on four different versions of the TT dataset: TT, TT-concat, TT-balanced, and TT-concat-balanced. TT-concat-balanced was constructed from TT-balanced (Table \ref{tab:demographic_info}) in the same way as TT-concat (Table \ref{tab:audio_stats}). For all TT datasets, we followed the same speaker-independent nested 10-fold cross-validation. The validation set is randomly selected from the training set and is not speaker-independent according to the official implementation.\footnotemark[7]

We used the same training configurations for all experiments as in the official implementation, with two exceptions: batch size and maximum length in seconds. Due to the limited memory of a single A10 GPU, we reduced the batch size to 16.
Given that the average audio length in the TT dataset is 1.4 seconds, we set the maximum length to 5 seconds to avoid excessive padding. For TT-concat and TT-concat-balanced, we kept the original 10-second limit. Apart from these two, the only difference across the experiments is the data. 

We found that retraining the model with different random seeds causes variable results, indicating that the training process is not entirely stable. To improve the reliability of the results, we trained each setting five times with different seeds and reported the average values. The standard deviation of the mean metrics across runs is around 1\% (with an overall range of about 2\%), which is consistent across all settings.

\section{Results and Discussion}\label{sec:results}

\begin{table*}[ht!]
\centering
\footnotesize
\vspace{-0.2cm}
\caption{This table compares the performance of PD classifiers across different training and test data settings. The mean±std of metrics (with multiplied by 100) were computed across 10 folds for each of the 5 runs and then averaged across runs.}
\vspace{-0.2cm}

\begin{tabular}{lllllll}
\Xhline{2\arrayrulewidth}
\textbf{Dataset} & \textbf{Accuracy} & \textbf{Precision} & \textbf{F1 Score} & \textbf{ROC-AUC} & \textbf{Sensitivity} & \textbf{Specificity} \\ \Xhline{2\arrayrulewidth}
PC-GITA~\cite{SSL4PR} & \textbf{82.21±8.2} & \textbf{83.43±7.8} & \textbf{81.99±8.3} & \textbf{88.42±8.4} & \textbf{75.22±12.3} & \textbf{89.21±9.2} \\ 
PC-GITA~\cite{SSL4PR} (Replicated) & 78.00±9.6 & 79.19±9.2 & 77.67±9.9 & 83.98±10.4 & 75.05±13.7 & 80.97±15.2 \\ \Xhline{2\arrayrulewidth}
TT & 69.17±11.6 & 69.24±11.9 & 68.67±11.8 & 72.92±14.4 & 71.29±13.5 & 66.66±14.2 \\ 
TT-concat & 70.65±13.1 & 70.91±13.9 & 69.69±14.1 & 79.02±13.8 & 72.63±12.3 & 68.13±22.8 \\ 
TT-balanced  & 71.10±8.0 & 73.03±8.2 & 70.42±8.3 & 80.24 ± 9.8 & 69.51±18.5 & 72.65±15.3 \\ 
\textbf{TT-concat-balanced}  & \textbf{74.70±10.4} & \textbf{76.33±10.1} & \textbf{74.19±10.8} & \textbf{84.24±10.7} & \textbf{71.91±16.5} & \textbf{77.50±15.8} \\ \Xhline{2\arrayrulewidth}
PC-GITA test on TT-concat-balanced & 52.91±4.4 & 52.68±25.2 & 39.63±8.3 & 74.38±14.7 & 98.96±2.5 & \underline{7.00}±9.5 \\ 
TT-concat-balanced test on PC-GITA  & 64.88±8.6 & 66.89±9.1 & 63.53±9.5 & 73.06±11.0 & 73.78±15.2 & 55.97±19.9 \\ \Xhline{2\arrayrulewidth}
\end{tabular}

\vspace{-0.35cm}
\label{tab:performance_comparison}
\end{table*}


\subsection{Experimental Results}
Table \ref{tab:performance_comparison} presents a performance comparison across different conditions.
We see that the F1 scores for PC-GITA (Replicated) are in the range of 0.75--0.80. 
If we take this range to be representative of the scores that would be needed to consider a classifier promising for further development, and ultimate use in a clinical setting, then we can see that the performances of TT datasets do not always land in this range.
We do see that performance on the TT-concat-balanced can touch the lower end of this range (\textit{RQ}1).
However, in order to reach this level, it is necessary to concatenate the TT data and also make sure that it is appropriately balanced (\textit{RQ}2).
We recommend that when using non-diagnostic speech data, researchers should not forget that these are key design decisions to make concerning how a dataset should be used and evaluated. 



We can imagine a scenario in which a model is trained on diagnostic data but applied to non-diagnostic data, and vice versa. As shown in the bottom two rows of Table \ref{tab:performance_comparison}, a model trained on PC-GITA generalizes poorly to TT-concat-balanced data, predicting nearly all cases as PD (F1=0.396, specificity=0.07). In contrast, a model trained on TT-concat-balanced achieves about 0.65 on most metrics when tested on PC-GITA (\textit{RQ}1). This asymmetry may stem from the different speech tasks used in the two datasets, with TT eliciting shorter phrases than PC-GITA, and could also be related to differences in TAD, with speech potentially being more affected in the PD patients from PC-GITA (\textit{RQ}2).
It is important to note that these two datasets are in different languages. Nevertheless, our findings indicate that further exploration of the generalizability of the PD speech dataset to other datasets holds promise.


\subsection{Failure Analysis}

\begin{figure}[ht!]
    \centering    \includegraphics[width=1.0\linewidth]{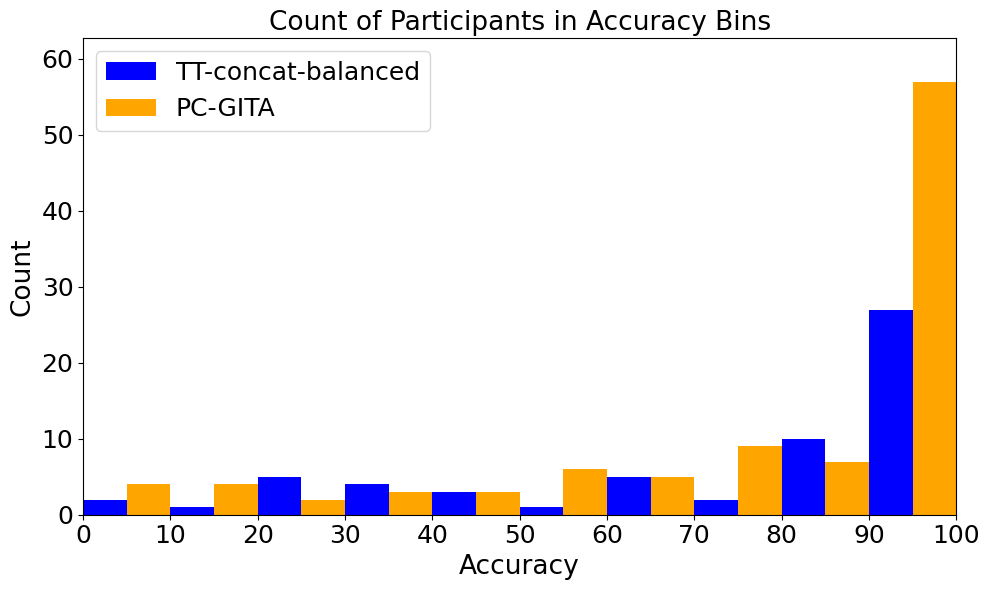}
    \vspace{-0.7cm}
    \caption{Count of participants in 10 accuracy bins.}
    \label{fig: acc_distribution}
        \end{figure}

To address \textit{RQ}3, we first evaluate the accuracy of each participants to identify the factors contributing to the large variability in classification performance across test folds. This is done by using the model trained on each fold's training set to evaluate the participant assigned to that fold's test set. The reported accuracy is averaged over five runs. Figure \ref{fig: acc_distribution} shows the distribution of speaker-level accuracy for TT-concat-balanced and PC-GITA, grouped into ten 10\% accuracy bins, representing the number of participants in each accuracy range. Both datasets follow a similar pattern: most participants achieve high accuracy (>90\%), but some have very low accuracy (<30\%).
This difference in accuracy among individual speakers explains the variability observed.

    \vspace{-0.15cm}

    \vspace{-0.2cm}

To further investigate the factors contributing to variability among individual speakers, we examined whether low speaker accuracy stemmed from random variation arising from the split generation. We generated a new split with a different seed again and retrained the model; no significant differences were observed. To explore whether the variability is due to the combination of multiple tasks, we trained and tested on TT model on only the BoundaryTone task of the TT.
Here, we also observed that performance was uneven across speakers. 

\begin{figure}[ht!]
            \vspace{-0.25cm}
    \centering    \includegraphics[width=1.0\linewidth]{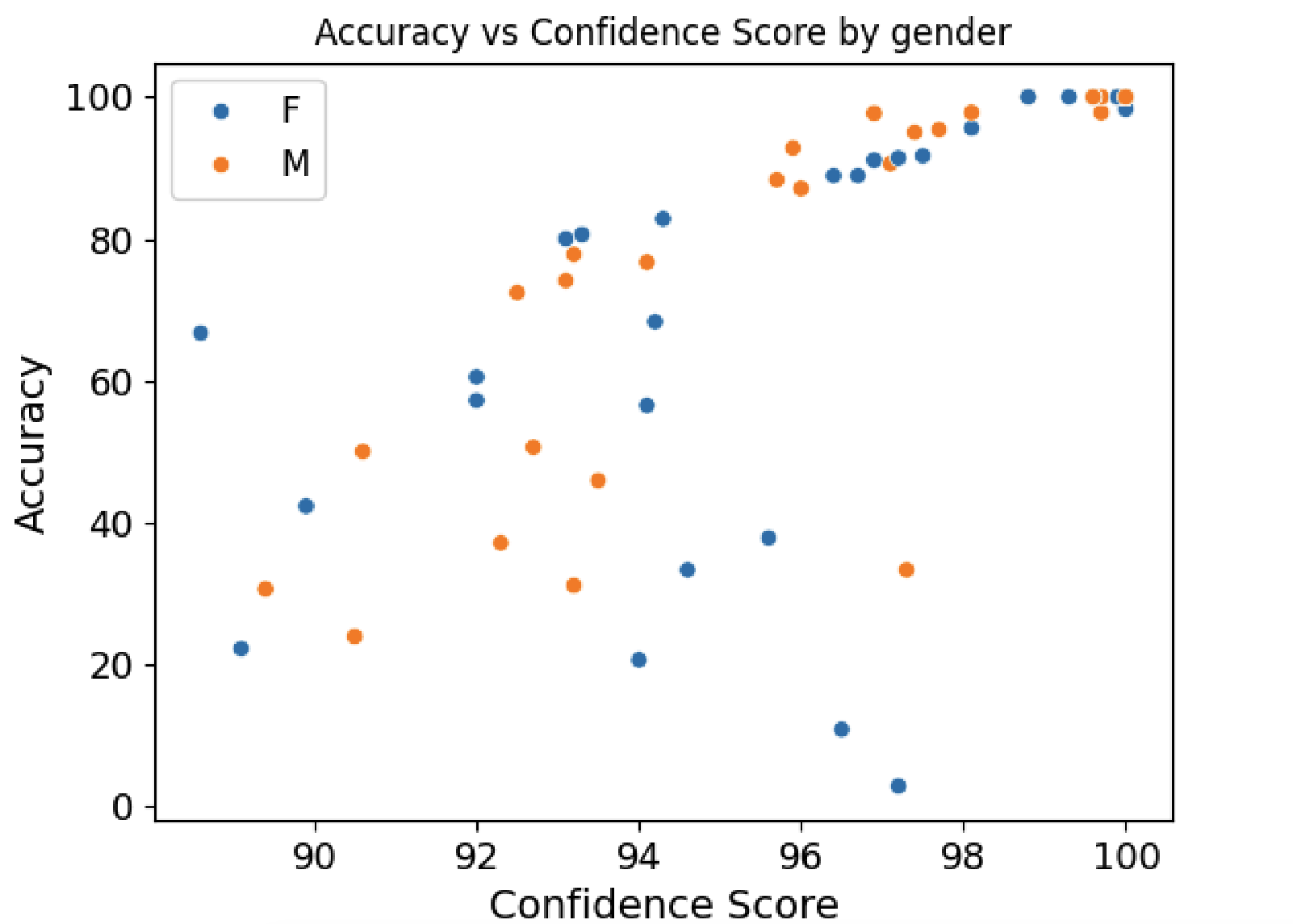}
        \vspace{-0.55cm}
    \caption{Accuracy vs CS, by gender, TT-concat-balanced.}
    \label{fig: acc_byCS}
    \vspace{-0.2cm}

        \end{figure}

We further analyzed the performance of individual participants from the TT-concat-balanced dataset. The three participants with the lowest accuracy (<20\%) are all female (2 PD, 1 HC), as shown in Figure \ref{fig: acc_byCS}. The average accuracy is 69.6\% for females compared to 82.1\% for males. However, the Mann-Whitney (M-W) U test~\cite{mcknight2010mann} (p=0.276) indicates that this difference is not statistically significant. No clear age-related pattern was found, as the differences in accuracy over different age ranges are minimal. 
We further analyzed by-speaker accuracy and found that across five runs, about 20 participants with very high accuracy (>95\%) consistently remained high, while only 4 participants with low accuracy (<50\%) consistently remained low. In particular, 11 participants exhibited accuracy differences exceeding 50\% across runs and had relatively low confidence scores. To explore whether specific speech features contributed to distinct performance patterns among these participants, we extracted 14 basic speech features from each participant's recordings, including jitter, shimmer, F0, energy, formants, and spectral features~\cite{zhao2007spectral}. We performed M-W U tests on the mean and standard deviation of these features for the top and bottom 20 participants based on average accuracy. However, no statistically significant differences were found for any of these features.

Finally, we calculated the Pearson correlation coefficient between accuracy per participant and three variables: TAD, ROMP speech score, and average confidence score (CS). The correlations were weak for TAD (0.259) and moderate for ROMP (0.374). This suggests a slight alignment between self-reported speech difficulties and the model's detection accuracy, where more severe self-reported speech problems correlate with better detection. A strong correlation was observed for CS (0.731), as shown in Figure \ref{fig: acc_byCS}, although average CS values remain consistently high (>90\%). This highlights the value of CS in improving prediction reliability and providing better support for clinical applications.


    


\section{Conclusion}
This study examined the feasibility of diagnosing PD using controlled setting speech data not originally intended for PD diagnostics. 
Our findings indicate that a dataset like TT can be as useful as traditional diagnostic datasets for developing a PD classifier while exhibiting even better generalization ability (\textit{RQ}1). This expands the range of data available for PD detection in the research community. Concatenating audio recordings and balancing participants' gender and disease status distributions affect classification scores within datasets (\textit{RQ}2). When the performance metrics varied substantially across test folds, this variability was attributed to differences in classification accuracy among individual speakers (\textit{RQ}3). However, the model's poor performance for specific participants remains unclear. We plan to incorporate explainable methods to better understand why the model makes incorrect decisions. Furthermore, involving clinical experts is essential to identify individual-specific perceptible features that may guide diagnosis in future research.

\section{Acknowledgments}
This publication is part of the project Responsible AI for Voice Diagnostics (RAIVD) with file number NGF.1607.22.013 of the research program NGF AiNed Fellowship Grants, which is financed by the Dutch Research Council (NWO). This work was conducted in close collaboration with the project “Turn-taking in Dialogue in Populations with Communicative Impairment" ({https://www.ru.nl/en/research/research-projects/turntaking-in-dialogue-in-populations-with-communicative-impairment}). This work used the Dutch national e-infrastructure with the support of the SURF Cooperative using grant no. EINF-10519. Finally, we extend our special thanks to the authors of the corresponding data materials for providing us with access.


\bibliographystyle{IEEEtran}

\bibliography{mybib}

\begin{thebibliography}{10}
\providecommand{\url}[1]{#1}
\csname url@samestyle\endcsname
\providecommand{\newblock}{\relax}
\providecommand{\bibinfo}[2]{#2}
\providecommand{\BIBentrySTDinterwordspacing}{\spaceskip=0pt\relax}
\providecommand{\BIBentryALTinterwordstretchfactor}{4}
\providecommand{\BIBentryALTinterwordspacing}{\spaceskip=\fontdimen2\font plus
\BIBentryALTinterwordstretchfactor\fontdimen3\font minus \fontdimen4\font\relax}
\providecommand{\BIBforeignlanguage}[2]{{%
\expandafter\ifx\csname l@#1\endcsname\relax
\typeout{** WARNING: IEEEtran.bst: No hyphenation pattern has been}%
\typeout{** loaded for the language `#1'. Using the pattern for}%
\typeout{** the default language instead.}%
\else
\language=\csname l@#1\endcsname
\fi
#2}}
\providecommand{\BIBdecl}{\relax}
\BIBdecl

\bibitem{1_ngo2022computerized}
Q.~C. Ngo, M.~A. Motin, N.~D. Pah, P.~Drot{\'a}r, P.~Kempster, and D.~Kumar, ``Computerized analysis of speech and voice for {P}arkinson's disease: A systematic review,'' \emph{Computer Methods and Programs in Biomedicine}, vol. 226, p. 107133, 2022.

\bibitem{poewe2017parkinson}
W.~Poewe, K.~Seppi, C.~M. Tanner, G.~M. Halliday, P.~Brundin, J.~Volkmann, A.-E. Schrag, and A.~E. Lang, ``Parkinson disease,'' \emph{Nature Reviews Disease Primers}, vol.~3, no.~1, pp. 1--21, 2017.

\bibitem{pdspeech1pinto2004treatments}
S.~Pinto, C.~Ozsancak, E.~Tripoliti, S.~Thobois, P.~Limousin-Dowsey, and P.~Auzou, ``Treatments for dysarthria in {P}arkinson's disease,'' \emph{The Lancet Neurology}, vol.~3, no.~9, pp. 547--556, 2004.

\bibitem{pdspeech2rusz2013imprecise}
J.~Rusz, R.~Cmejla, T.~Tykalova, H.~Ruzickova, J.~Klempir, V.~Majerova, J.~Picmausova, J.~Roth, and E.~Ruzicka, ``Imprecise vowel articulation as a potential early marker of {P}arkinson's disease: effect of speaking task,'' \emph{The Journal of the Acoustical Society of America}, vol. 134, no.~3, pp. 2171--2181, 2013.

\bibitem{mu2017parkinson}
J.~Mu, K.~R. Chaudhuri, C.~Bielza, J.~de~Pedro-Cuesta, P.~Larra{\~n}aga, and P.~Martinez-Martin, ``Parkinson's disease subtypes identified from cluster analysis of motor and non-motor symptoms,'' \emph{Frontiers in Aging Neuroscience}, vol.~9, p. 301, 2017.

\bibitem{biomarkertelho2024speech}
C.~Botelho, A.~Abad, T.~Schultz, and I.~Trancoso, ``Speech as a biomarker for disease detection,'' \emph{IEEE Access}, vol.~12, pp. 184\,487--184\,508, 2024.

\bibitem{SSL4PR}
M.~{La Quatra}, M.~F. Turco, T.~Svendsen, G.~Salvi, J.~R. Orozco-Arroyave, and S.~M. Siniscalchi, ``Exploiting foundation models and speech enhancement for {P}arkinson's disease detection from speech in real-world operative conditions,'' in \emph{Interspeech 2024}, 2024, pp. 1405--1409.

\bibitem{Foundation4PDpurohitautomatic}
T.~Purohit, B.~Ruvolo, J.~R. Orozco-Arroyave, and M.~Magimai.-Doss, ``Automatic parkinson’s disease detection from speech: Layer selection vs adaptation of foundation models,'' in \emph{ICASSP 2025 - 2025 IEEE International Conference on Acoustics, Speech and Signal Processing (ICASSP)}, 2025, pp. 1--5.

\bibitem{PCGITA2oliveiraPilotStudySpeech2025}
G.~C. Oliveira, N.~D. Pah, Q.~C. Ngo, A.~Yoshida, N.~B. Gomes, J.~P. Papa, and D.~Kumar, ``A pilot study for speech assessment to detect the severity of {{Parkinson}}'s disease: {{An}} ensemble approach,'' \emph{Computers in Biology and Medicine}, vol. 185, p. 109565, 2025.

\bibitem{vasquez-correaConvolutionalNeuralNetworks2019}
J.~C. V{\'a}squez-Correa, T.~Arias-Vergara, C.~D. Rios-Urrego, M.~Schuster, J.~Rusz, J.~R. Orozco-Arroyave, and E.~N{\"o}th, ``Convolutional neural networks and a transfer learning strategy to classify {P}arkinson’s disease from speech in three different languages,'' in \emph{Progress in Pattern Recognition, Image Analysis, Computer Vision, and Applications: 24th Iberoamerican Congress, CIARP 2019, Havana, Cuba, October 28-31, 2019, Proceedings 24}.\hskip 1em plus 0.5em minus 0.4em\relax Springer, 2019, pp. 697--706.

\bibitem{referralsprabhod2024role}
K.~J. Prabhod, ``The role of artificial intelligence in reducing healthcare costs and improving operational efficiency,'' \emph{Quarterly Journal of Emerging Technologies and Innovations}, vol.~9, no.~2, pp. 47--59, 2024.

\bibitem{observerSVENSBERG20212136}
K.~Svensberg, B.~G. Kalleberg, L.~Mathiesen, Y.~Andersson, S.~E. Rognan, and S.~K. Sporrong, ``The observer effect in a hospital setting - experiences from the observed and the observers,'' \emph{Research in Social and Administrative Pharmacy}, vol.~17, no.~12, pp. 2136--2144, 2021.

\bibitem{ParkCeleb_favaroUnveilingEarlySigns2024}
A.~Favaro, A.~Butala, T.~Thebaud, J.~Villalba, N.~Dehak, and L.~Moro-Velázquez, ``Unveiling early signs of {{Parkinson}}'s disease via a longitudinal analysis of celebrity speech recordings,'' \emph{npj Parkinsons Dis.}, vol.~10, no.~1, p. 207, 2024.

\bibitem{wiemann2017turn}
J.~M. Wiemann and M.~L. Knapp, ``Turn-taking in conversations,'' \emph{Communication Theory}, pp. 226--245, 2017.

\bibitem{LisannevangelderenInnovativeSpeechBasedDeep2024}
L.~van Gelderen and C.~Tejedor-García, ``Innovative speech-based deep learning approaches for {P}arkinson’s disease classification: A systematic review,'' \emph{Applied Sciences}, vol.~14, no.~17, 2024.

\bibitem{PCGITAorozco2014new}
J.~R. Orozco-Arroyave, J.~D. Arias-Londo{\~n}o, J.~F. Vargas-Bonilla, M.~C. Gonz{\'a}lez-R{\'a}tiva, and E.~N{\"o}th, ``New {S}panish speech corpus database for the analysis of people suffering from {P}arkinson`s disease,'' in \emph{Proceedings of the Ninth International Conference on Language Resources and Evaluation ({LREC}`14)}, 2014, pp. 342--347.

\bibitem{Neurovozmendes2024neurovoz}
J.~Mendes-Laureano, J.~A. G{\'o}mez-Garc{\'\i}a, A.~Guerrero-L{\'o}pez, E.~Luque-Buzo, J.~D. Arias-Londo{\~n}o, F.~J. Grandas-P{\'e}rez, and J.~I. Godino-Llorente, ``Neuro{V}oz: a {Castilian Spanish} corpus of parkinsonian speech,'' \emph{Scientific Data}, vol.~11, no.~1, p. 1367, 2024.

\bibitem{pictureglaser1992picture}
W.~R. Glaser, ``Picture naming,'' \emph{Cognition}, vol.~42, no. 1-3, pp. 61--105, 1992.

\bibitem{earlylatebogels2018planning}
S.~B{\"o}gels, M.~Casillas, and S.~C. Levinson, ``Planning versus comprehension in turn-taking: Fast responders show reduced anticipatory processing of the question,'' \emph{Neuropsychologia}, vol. 109, pp. 295--310, 2018.

\bibitem{boundarytonebarthel2017next}
M.~Barthel, A.~S. Meyer, and S.~C. Levinson, ``Next speakers plan their turn early and speak after turn-final “go-signals”,'' \emph{Frontiers in Psychology}, vol.~8, p. 393, 2017.

\bibitem{PCGITA3gallo-aristizabalAutomaticClassificationParkinsons2024a}
J.~D. Gallo-Aristizábal, D.~Escobar-Grisales, C.~D. Ríos-Urrego, E.~Nöth, and J.~R. Orozco-Arroyave, ``Automatic classification of {P}arkinson’s disease using wav2vec embeddings at phoneme, syllable, and word levels,'' in \emph{Text, {{Speech}}, and {{Dialogue}}}, 2024, vol. 15049, pp. 313--323.

\bibitem{Rompkalf2011reproducibility}
J.~G. Kalf, G.~F. Borm, B.~J. de~Swart, B.~R. Bloem, M.~J. Zwarts, and M.~Munneke, ``Reproducibility and validity of patient-rated assessment of speech, swallowing, and saliva control in {P}arkinson's disease,'' \emph{Archives of Physical Medicine and Rehabilitation}, vol.~92, no.~7, pp. 1152--1158, 2011.

\bibitem{gimeno2025unveiling}
D.~Gimeno-Gómez, C.~Botelho, A.~Pompili, A.~Abad, and C.-D. Martínez-Hinarejos, ``Unveiling interpretability in self-supervised speech representations for parkinson’s diagnosis,'' \emph{IEEE Journal of Selected Topics in Signal Processing}, p. 1–14, 2025.

\bibitem{chen2022wavlm}
S.~Chen, C.~Wang, Z.~Chen, Y.~Wu, S.~Liu, Z.~Chen, J.~Li, N.~Kanda, T.~Yoshioka, X.~Xiao \emph{et~al.}, ``Wavlm: Large-scale self-supervised pre-training for full stack speech processing,'' \emph{IEEE Journal of Selected Topics in Signal Processing}, vol.~16, no.~6, pp. 1505--1518, 2022.

\bibitem{mcknight2010mann}
P.~E. McKnight and J.~Najab, ``{Mann-Whitney U Test},'' \emph{The Corsini Encyclopedia of Psychology}, pp. 1--1, 2010.

\bibitem{zhao2007spectral}
Z.~Zhao and H.~Liu, ``Spectral feature selection for supervised and unsupervised learning,'' in \emph{Proceedings of the 24th International Conference on Machine Learning}, 2007, pp. 1151--1157.

\end{thebibliography}

\end{document}